\documentclass[aps,prl,preprint,superscriptaddress]{revtex4}

\begin{document}

\title{Long-Range Order of Vortex Lattices Pinned 
by Point Defects in Layered Superconductors}

\author{J. P. Rodriguez}
\thanks{Permanent address: Department of Physics and Astronomy,
California State University, Los Angeles, California 90032}

\affiliation {Superconductivity Technology Center,
Los Alamos National  Laboratory, Los Alamos, NM 87545}

\date{\today}

\begin{abstract}
How  the   vortex lattice 
orders at long range in a layered superconductor
with weak point  pinning centers
is studied through a duality analysis of the corresponding
frustrated $XY$ model.
Vortex-glass order emerges out of the vortex liquid across a macroscopic number of
weakly coupled  layers
in perpendicular magnetic field as the system cools down.
Further, the naive magnetic-field scale determined by the Josephson coupling
between adjacent layers is found to serve as an upperbound for the
stability of any possible 
conventional vortex lattice
phase at low temperature
in the extreme type-II limit.
\end{abstract}

\maketitle

\section {Introduction}
It is well known that an
external magnetic field
can penetrate a
type-II superconductors
in the form of lines  of flux quanta\cite{Tink}.
The repulsive forces that such flux lines experience favor the
creation of  a triangular vortex lattice,
while the quenched point disorder present to some degree in
all superconductors frustrates that tendency.
Three thermodynamic groundstates are then likely.
Either the triangular vortex lattice is robust to weak point
pinning and assumes  a Bragg glass state 
with no lines of dislocations that thread it\cite{giamarchi},
or it will transit into a defective state
with quenched-in lines of dislocations that thread it.
The latter, in turn, has two possible outcomes: 
a vortex glass state that
retains macroscopic phase coherence
of the superconducting order parameter\cite{ffh},
or a pinned liquid state that does not\cite{giamarchi}.

High-temperature superconductors, in particular, 
are extremely type-II and layered\cite{Tink}.
Below, we shall study 
how a   vortex lattice pinned by material point defects
orders at long-range in such materials.
The vortex lattice in layered superconductors with weak random point pins
shall be described theoretically
in terms of the phase  of the superconducting order parameter
via the corresponding frustrated $XY$ model\cite{nono1}.
This model notably neglects the effects of magnetic coupling between layers,
while it treats the Josephson coupling between them   exactly.
The growth of long-range order across layers is then computed from the
$XY$ model through a duality analysis\cite{jpr00},
where the ratio of the energy of the Josephson coupling 
between adjacent layers to the temperature emerges as a small parameter.
We find first that the correlation length
for 
vortex-glass order\cite{ffh}
across weakly coupled layers diverges
as   temperature cools down
from the vortex liquid
towards the two-dimensional (2D) ordering transition.
The divergence signals a transition to 
a vortex glass phase\cite{olsson}\cite{bulaevskii}\cite{jpr04a}\cite{zeldov05}.
Second, 
we find no evidence for the divergence of conventional superconducting
phase correlations across layers
from inside the latter vortex glass 
to lowest order in the  inter-layer Josephson coupling.
This indicates ultimately that the naive decoupling field for the
pristine vortex lattice\cite{vinokur} serves as an
upper bound for a stable Bragg glass phase\cite{giamarchi}
in the extreme type-II limit.
Comparisons with  previous numerical\cite{nono1}, theoretical\cite{K-V},
and experimental\cite{neutrons} determinations
of the stability line for the Bragg glass in layered superconductors
are made at the end of the paper.

\section {Two Dimensions}
The $XY$ model with uniform frustration is the minimum
theoretical description of  vortex matter in
extremely type-II 
superconductors.  Both fluctuations of the magnetic induction and
of the magnitude of the superconducting order parameter are neglected
within this approximation.
The model hence  is valid deep inside
the interior of the mixed phase.  
The thermodynamics of an isolated  layer
with uniform frustration
is determined by its superfluid kinetic energy
\begin{equation}
E_{XY}^{(2)} = -\sum_{\vec r}\sum_{\mu=x,y} J_{\mu} {\rm cos}
[\Delta_{\mu}\phi  - A_{\mu}]|_{\vec r},
\label{2dxy}
\end{equation}
which is a  functional of the phase 
of the superconducting order parameter,
$e^{i \phi}$,
over the square lattice, $\vec r$.
Here, $J_x$  and   $J_y$
are the local phase rigidities that are 
equal and constant,
except over links in the vicinity
of a pinning center.
The   vector potential
$\vec A = (0, 2\pi f x/a)$
represents the magnetic induction
oriented perpendicular to the layers,
$B_{\perp} = \Phi_0 f / a^2$.
Here $a$ denotes the square lattice constant, which is of order
the coherence length of the Cooper pairs, $\Phi_0$ denotes
the flux quantum, and $f$ denotes the concentration of vortices per site.

Analytical and numerical work indicates that
the 2D vortex lattice is invaded by
quenched-in dislocations in the presence of any degree of
random point pinning\cite{n-s}.
The author has argued\cite{jpr04c} that the dislocations quenched into
each 2D vortex lattice described by the frustrated $XY$ model (\ref{2dxy})
notably do {\it not} line up to form low-angle grain boundaries, however 
(cf. ref. \cite{chandran}).
That argument is based on
the incompressible nature of 2D vortex matter
in the extreme type-II limit.
The absence of grain boundaries
is consistent with
Monte Carlo simulations\cite{jpr-cc} of the equivalent
2D Coulomb gas ensemble with random point pins\cite{jpr04a},
as well as 
with Monte Carlo simulations of the frustrated $XY$ model
in three dimensions
with randomly located columnar pins\cite{nono2}.
Secondly, a net superfluid density 
is predicted at zero temperature for perpendicular magnetic fields above the
collective-pinning threshold, $B_{\rm cp}^{(2D)}$,
in which case
the number of pinned vortices is greater than
the number of
isolated dislocations quenched into the 2D vortex lattice\cite{jpr05}.
Here, the scale of the Larkin domains\cite{Tink}
is set by the separation between neighboring dislocations
quenched into the vortex lattice.
A variational calculation by Mullock and Evetts
yields the estimate
$B_{\rm cp}^{(2D)}\sim (4 f_{\rm p}/ \varepsilon_0 d)^2\Phi_0$
for the threshold field\cite{M-E}, 
where $f_p$ denotes the maximum pinning force,
where
$\varepsilon_0 = (\Phi_0/4\pi \lambda_L)^2$ 
is the maximum tension of  a fluxline in the superconductor,
and where $d$ denotes the separation between adjacent layers.
Here $\lambda_L$ represents the London penetration depth.  	
The  pinning of the vortex lattice in isolated layers
shall be assumed to be
collective henceforth: $B_{\perp} > B_{\rm cp}^{(2D)}$.

The previous indicates that
a hexatic vortex glass characterized by 
a homogeneous distribution of quenched-in dislocations
and by  a net superfluid density
exists in isolated layers of the frustrated $XY$ model (\ref{2dxy})
with weak random point pins  at zero temperature\cite{jpr05}.
The transition temperature $T_g^{(2D)}$
that separates the low-temperature hexatic vortex glass
from the high-temperature vortex liquid must  therefore
be  equal to zero or greater.
Recent current-voltage measurements of 2D arrays of
Josephson junctions in weak external magnetic field indicate
that the 2D superconducting/normal transition at $T = T_g^{(2D)}$
is second order\cite{arrays}, with $T_g^{(2D)}$ much
larger than the
2D melting temperature of  the pristine vortex lattice,
$T_m^{(2D)} \cong J/20$. 
Since the previous is a faithful realization of 
the frustrated $XY$ model (\ref{2dxy}) in 2D
with random point pinning centers, 
we shall assume henceforth that the hexatic vortex glass melts   
into a vortex liquid at temperature $T_g^{(2D)} > 0$ via 
a second-order phase transition.

\section {Three Dimensions}
We shall now demonstrate how long-range
vortex-glass order emerges across layers
from the vortex liquid phase of layered superconductors
with weak random point pins.
Let us first couple the layers through the Josephson effect
by adding a term
$-J_z {\rm cos} (\Delta_z \phi - A_z)$
to the internal energy of the frustrated $XY$ model
for each nearest-neighbor
link across adjacent layers.
The component of the magnetic induction parallel to the
layers is taken to be null throughout.
At weak coupling, $J_z\ll k_B T$,
phase correlations across $N$ layers
can then be determined
from the quotient
\begin{equation}
  \Bigl\langle {\rm exp} \Bigl[i\sum_r p(r) \phi(r)\Bigr]\Bigr\rangle =
Z_{\rm CG}[p]/Z_{\rm CG}[0]
\label{quo}
\end{equation}
of partition functions for
a    layered Coulomb gas (CG) ensemble\cite{jpr00}:
\begin{equation}
  Z_{\rm CG}[p] = \sum_{\{n_{z}(r)\}} y_0^{N[n_z]}
\cdot
\Pi_{l} C_l [q_l]
\cdot e^{-i\sum_r n_z A_z}.
\label{z_cg}
\end{equation}
Above,
$n_z (\vec r, l)$ is a  dual charge/integer field
that lives on links between adjacent layers $l$ and $l+1$,
located  at 2D points $\vec r$,
and $p(r) = \delta_{\vec r, 0} \cdot (\delta_{l, 1} - \delta_{l, N})$
is the  external   integer probe  field.
The ensemble is weighted
by a product
of phase auto-correlation functions
for isolated layers $l$,
\begin{equation}
C_l [q] = \langle {\rm exp} 
[i\sum_{\vec r} q (\vec r) \phi (\vec r, l)] \rangle_{J_z = 0},
\label{c_l}
\end{equation}
probed at the dual   charge  that accumulates onto
that layer:
\begin{equation}
 q_l (\vec r) = p(\vec r, l) +  n_z (\vec r, l-1) - n_z (\vec r, l).
\label{ql}
\end{equation}
It is also weighted
by a bare fugacity
$y_0$   that is
raised to the power
$N [n_z]$
equal to the total
 number of dual charges, $n_z = \pm 1$.
The fugacity of the dual CG ensemble (\ref{z_cg})
 is given by
$y_0 = J_z / 2 k_B T$ in the selective high-temperature regime,
$J_z \ll k_B T$,
reached at large model  anisotropy.
It is small compared to unity in such case, which implies a dilute
concentration of dual $n_z$ charges\cite{jpr00}.
The dual CG ensemble (\ref{z_cg})
is valid in that  regime.

The above duality analysis is particularly natural and effective
in the vortex-liquid phase,
where autocorrelations of the superconducting order parameter
in isolated layers (\ref{c_l}) are short range.
They shall be assumed to take to the form
that is characteristic of a hexatic vortex liquid
between points $\vec r_1$ and $\vec r_2$ 
in an isolated  layer $l$\cite{jpr05}\cite{jpr01}:
\begin{equation}
C_l (1, 2) = g_0 e^{- r_{1, 2} / \xi_{2D}} 
e^{-i\phi_0 (1)} e^{i\phi_0 (2)}.
\label{form1}
\end{equation}
Here  $e^{i\phi_0}$ 
is the superconducting
order parameter of layer $l$ in isolation at zero temperature,
$\xi_{2D}$ denotes the phase correlation length of
 the 2D hexatic vortex liquid,
and $g_0$ is a prefactor of order unity.
Also, $\vec r_{1,2} = \vec r_1 - \vec r_2$ is the displacement
between the probes within layer $l$.
To lowest order in the  (dual) fugacity, $y_0$, 
Eqs. (\ref{quo}) and (\ref{z_cg})
then yield the expression
\begin{equation}
\overline{\langle e^{i\phi_{l, l+n}}\rangle} \cong 
y_0^n \sum_1 ... \sum_n \overline{
C_l (0,1) \cdot C_{l+1} (1,2)\cdot  ... \cdot C_{l+n} (n,0)}
\label{auto1}
\end{equation}
for the bulk average (overbar)
of  the gauge-invariant auto-correlation function
of  the conventional superconducting order parameter
$e^{i\phi}$  across $n$ layers,
at zero parallel field.
Above and hereafter, 
we take the gauge $A_z = 0$.
The uncorrelated nature of point pinning centers across layers
implies the form
\begin{equation}
\overline{\Pi_{m=0}^n 
e^{-i\phi_0 (\vec r_m, l+m)} e^{i\phi_0 (\vec r_{m+1}, l+m)}}
= \Pi_{m=0}^n e^{-r_{m, m+1}/2 l_{\phi}}
\label{form2}
\end{equation}
%
for the bulk average of the relevant
product of zero-temperature order parameters,
with matching endpoints $\vec r_0 = \vec r_{n+1}$.
Here, $l_{\phi}$ is a
quenched  disorder scale
that is set by the density of lines  of dislocations
quenched  into the vortex lattice
at $J_z = 0$
that begin or  end at a given layer.
We remind the reader that $l_{\phi}$ is believed to be finite
(in the absence of inter-layer coupling)
for any non-zero strength of quenched point disorder\cite{n-s}. 
Substitution of (\ref{form2}) into expression (\ref{auto1})
then yields the principal dependence\cite{integral}
\begin{equation}
\overline{\langle e^{i\phi_{l, l+n}}\rangle} \propto 
[g_0 (J / k_B T) ((l_{\phi}^{-1} + \xi_{\phi}^{-1})^{-1} /  \Lambda_0)^2]^n
\label{result1}
\end{equation}
for the correlation of the 
conventional superconducting order parameter across $n$ layers.
Here, $J$ is the macroscopic phase rigidity of 
an isolated layers at zero temperature,
$\Lambda_0 =  (J/J_z)^{1/2} a$ is the Josephson penetration depth,
and $\xi_{\phi} = \xi_{2D} / 2$.
Notice that the  existence
of the disorder scale $l_{\phi}$
implies that the perturbative result  (\ref{result1}) above
does {\it not} diverge with 
the 2D phase correlation length $\xi_{2D}$
in the vicinity  of the 2D ordered phase.
We conclude that conventional superconducting phase coherence 
across many layers ($n\rightarrow\infty$) 
does not emerge out of the vortex liquid
at weak Josephson coupling between adjacent layers.

The growth of macroscopic  vortex-glass order
across layers\cite{ffh}
from inside of  the vortex liquid 
is still possible, however.
We shall test for it by computing the corresponding
auto-correlation function\cite{ffh},
which is given by
\begin{equation}
\overline{|\langle e^{i\phi_{l, l+n}}\rangle|^2} \cong 
y_0^{2 n} \sum_{1, \bar 1}  ...  \sum_{n, \bar n} 
\overline{C_l (0,1) C_l^* (0, \bar 1) 
\cdot C_{l+1}(1,2) C_{l+1}^*  (\bar 1, \bar 2) \cdot ...
\cdot C_{l+n}(n,0) C_{l+n}^* (\bar n, 0)}
\label{auto2}
\end{equation}
to lowest order in the (dual) fugacity, $y_0$.
It is natural to  look for vortex-glass order to emerge 
from within 
the 2D critical regime: 
$\xi_{2D} \gg  2 l_{\phi}$ at $T > T_g^{(2D)}$,
where $T_g^{(2D)}$ denotes the transition temperature of the 2D
hexatic vortex glass.
The bulk average of the product of
zero-temperature order parameters that appears in the integrand above
can then be approximated
by the corresponding  product of the bulk averages limited 
to adjacent layers, $l^{\prime} = l+m-1$ and $l^{\prime} + 1$, only:
%
\begin{equation}
\overline{{\rm exp} [{i\phi_{l^{\prime}, l^{\prime} + 1}^{(0)} (m)}]
\cdot 
{\rm exp} [{-i\phi_{l^{\prime}, l^{\prime} + 1}^{(0)} (\bar m)}]}
=
 e^{-r_{m, \bar m}/l_{\phi}}.
\label{form3}
\end{equation}
Here
$\phi_{l^{\prime}, l^{\prime}+1}^{(0)} (\vec r) = 
\phi_0 (\vec r,  l^{\prime} + 1)
 - \phi_0 (\vec r,  l^{\prime}) - A_z (\vec r)$
is the quenched inter-layer phase difference. 
Converting to center-of-mass variables among the inter-layer
coordinates, $\vec r_{m}$ and $\vec r_{\bar m}$,
then yields the principal dependence\cite{integral}
\begin{equation}
\overline{|\langle e^{i\phi_{l, l+n}}\rangle|^2} \propto
[g_0 (J / k_B T) (l_{\phi} \xi_{\phi} / \Lambda_0^2)]^{2 n}
\label{result2}
\end{equation}
for the vortex-glass correlations across layers 
in  the 2D critical regime, $\xi_{2D} \gg 2 l_{\phi}$, at zero parallel field.
The  corresponding correlation length ($\xi_{\perp}$)
is equal to the layer spacing ($d$)
when the argument in brackets above is set to $1/e$.
This occurs at a cross-over field
\begin{equation}
B_{\times} \sim
g_0 (J/k_B T) (l_{\phi} \xi_{\phi}/a_{\rm vx}^2) (\Phi_0 / \Lambda_0^2)
\label{xover}
\end{equation}
that separates two-dimensional
from three-dimensional (3D) vortex-liquid behavior 
(see Table \ref{mesa}).
Above, $a_{\rm vx}$ denotes the square root of the area per vortex
inside of a given layer.
Also, the argument between brackets on the right-hand side of
Eq. (\ref{result2}) notably diverges with $\xi_{2D}$
in the vicinity of the 2D ordering transition.
This indicates that a transition to a vortex glass 
that orders across a macroscopic number of layers\cite{nono1}\cite{olsson},
$\xi_{\perp}\rightarrow\infty$,
occurs at a  critical temperature $T_g$ that lies inside of the
window $[T_g^{(2D)}, T_{\times}]$.
Indeed, setting the argument of the exponent on the right-hand
side of Eq. (\ref{result2}) to unity yields a critical field
$B_g = B_{\times} / e$, below which a vortex glass exists
(see Table \ref{mesa}).

Last, recall that the superfluid density across layers,
$\rho_s^{\perp} = 
-{\cal N}^{-1} k_B T \partial ^2 {\rm ln} Z_{\rm CG} / \partial A_z^2|_0$,
is given by the expression\cite{jpr00}
\begin{equation}
\rho_s^{\perp}
= {\cal N}^{-1}
\Bigl \langle \Bigl [\sum_{\vec r, l} n_z (\vec r, l)\Bigr]^2 \Bigr\rangle
  k_B T,
\label{rhoperp}
\end{equation}
where ${\cal N}$ counts the number of nearest-neighbor links between
layers, and where periodic boundary conditions are assumed 
across layers.  
Study of Eqs. (\ref{quo})-(\ref{ql}) yields that
the tension for a line across layers  
of dual $n_z$ quanta
is equal to $\xi_{\perp}^{-1}$,
where $\xi_{\perp}$ denotes the correlation length
for vortex-glass order
across layers.
 The corresponding superfluid density
(\ref{rhoperp}) is then null 
in the limit of a macroscopic number of layers
inside of the vortex liquid,
where $\xi_{\perp} < \infty$
 (see Table \ref{mesa}).

The previous result (\ref{result2})
clearly demonstrates that a selective high-temperature
expansion in powers of the fugacity $y_0$ necessarily
breaks down in the 2D ordered phase, $T \leq T_g^{(2D)}$, 
where $\xi_{2D}$ is infinite. 
A direct analysis of the frustrated $XY$ model  for
an isolated layer finds,
in particular,
that long-range correlations 
of the superconducting order parameter
(\ref{c_l}) decay algebraicly instead
at such low temperatures\cite{jpr04a}\cite{jpr01}:
%
\begin{equation}
C_l [q]
    = g_0^{n_+}\cdot {\rm exp}\Bigl[  \eta_{2D}
\sum_{(1,2)} q(\vec r_1){\rm ln} (r_{1,2} / r_0)\, q(\vec r_2)\Bigr] \cdot
{\rm exp} \Bigl[i\sum_{1} q(\vec r_1)  \phi_0(\vec r_1, l)\Bigr].
\label{form4}
\end{equation}
%
The exponent
$\eta_{2D}$ that characterizes the algebraic decay of 2D phase coherence
is related to the 2D superfluid density
by
$\rho_s^{(2D)} = k_B T / 2\pi \eta_{2D}$.
Above,
$g_0 = \rho_s^{(2D)} / J$ is the ratio of the 2D phase stiffness
with its value at zero temperature, $J$,
while  $n_+$ counts half the number of probes in $q (\vec r)$.
Also, $r_0$ denotes the natural ultraviolet scale.
It is important
to observe 
at this stage
that the
loop excitations in the  (completely) dual
representation of the 3D XY model\cite{jpr00}
lose their integrity in the ordered phase.
This translates into the absence of charge conservation in the
(partially) dual CG ensemble (\ref{z_cg}).
In other words,
the dual $n_z$ charges
form a plasma in the ordered phase.
A Hubbard-Stratonovich transformation
of the CG partition function (\ref{z_cg})
followed by
the unrestricted summation of configurations of charges
with values $n_z = 0, \pm 1$
then yields
the equivalent  partition function\cite{jpr97}
$Z_{\rm LD}[p] = \int {\cal D} \theta\,  e^{-E_{\rm LD}/k_B T 
+ i\sum p\cdot \theta}$
for a   renormalized Lawrence-Doniach (LD) model
that shows no explicit dependence on the perpendicular magnetic field.
Its energy functional is specifically given by\cite{jpr00}\cite{jpr04a}
\begin{equation}
E_{\rm LD} = 
\rho_s^{(2D)}\int d^2 r
\sum_{l} \Biggl[
{1\over 2}(\vec\nabla\theta_l)^2
-\Lambda_0^{-2}
{\rm cos}\, \theta_{l, l+1} \Biggr],
\label{e_ld}
\end{equation}
where 
$\theta_{l, l+1}  = \phi_{l, l+1}^{(0)} +  \theta_{l+1} - \theta_l$.
The above continuum
description  is understood to have an  ultraviolet
cut off $r_0$  of order the inter-vortex spacing $a_{\rm vx}$.

We can now determine the growth
of correlations across
layers of the  conventional
superconducting  order parameter
deep inside of the  vortex glass phase, $T < T_g^{(2D)}$,
at weak Josephson coupling between layers, $\Lambda_0\rightarrow\infty$. 
The physics described by the original layered $XY$ model coincides directly
with that of the renormalized LD model described above
at large scales in distance compared to the ultraviolet cutoff, $r_0$.
Asymptotic correlations
of the conventional superconducting order parameter across layers, 
for example,
are identical to those of the LD model:  
${\rm lim}_{n\rightarrow\infty}
\overline{\langle e^{i\phi_{l, l+n}}\rangle} = 
\overline{\langle e^{i\theta_{l, l+n}}\rangle}$.
The configuration that optimizes $E_{\rm LD}$
must be determined first in order to compute the later
near zero temperature.
The LD energy functional  (\ref{e_ld}) implies that it satisfies the
field equation 
\begin{equation}
-\nabla^2 [\theta_{l^{\prime}+1}^{(0)} - \theta_{l^{\prime}}^{(0)}] = 
 \Lambda_0^{-2}{\rm sin}\, \theta_{l^{\prime}+1, l^{\prime}+2}^{(0)}
-2 \Lambda_1^{-2} {\rm sin}\, \theta_{l^{\prime}, l^{\prime}+1}^{(0)}
 +\Lambda_0^{-2} {\rm sin}\, \theta_{l^{\prime}-1, l^{\prime}}^{(0)},
\label{fldeq}
\end{equation}
where $\Lambda_1 = \Lambda_0$ (cf. refs. \cite{B-C} and \cite{magnetic}).
The phase angles  $\theta_{l^{\prime}}^{(0)}$ 
are  then constant inside of a given layer $l^{\prime}$
in the weak coupling limit\cite{jpr04a},
$\Lambda_0,\Lambda_1\rightarrow\infty$.
Next, if $\delta\theta_{l^{\prime}}^{(0)}$ denotes the fluctuation in 
the phase angles,
the auto-correlation function for conventional superconducting
order across many layers
is then approximated by the expression
\begin{equation}
\overline{ e^{i\theta_{l, l+n}}} \cong
\overline{\Pi_{l^\prime = l}^{l+n-1}
e^{i\theta_{l^{\prime}, l^{\prime}+1}^{(0)}}
\cdot i[\delta\theta_{l^{\prime}+1}^{(0)} - \delta\theta_{l^{\prime}}^{(0)}]}
\label{product}
\end{equation}
near zero temperature,
to lowest order in the fluctuation\cite{larkin}. 
After inverting the field equation
(\ref{fldeq}) for the fluctuation of the phase difference between
adjacent layers, 
substitution into the expression above yields the result
\begin{equation}
\overline{e^{i\theta_{l, l+n}}} \cong
a_n \Lambda_1^{-2n}
\Bigl[\Pi_{m=1}^n \int d^2 r_m G^{(2)}(0,m)\Bigr]
\overline{\Pi_{m=1}^n
e^{i\phi_{\l+m-1, l+m}^{(0)}(0)} e^{-i\phi_{\l+m-1, l+m}^{(0)}(m)}}
\label{auto3}
\end{equation}
%
for the autocorrelation
of  the superconducting order parameter across layers.
The prefactor on the right-hand side satisfies the recursion relation
$a_{n+1} = a_n + (\Lambda_1^2 / 2\Lambda_0^2)^2 a_{n-1}$,
with $a_{0} = 1$ and $a_{-1} = 0$.  Also, 
\begin{equation}
G^{(2)} = 
[-\nabla^2 + 
2 \Lambda_1^{-2} {\rm cos}\, \theta_{l^{\prime}, l^{\prime}+1}^{(0)}]^{-1}
\label{g2d}
\end{equation}
is the 2D Greens function.
The eigenstates of the latter operator within brackets are localized,
with a localization length\cite{russ}
$R_0\sim \Lambda_1^2/l_{\phi}$.
We therefore have
$G^{(2)} (1,2) = (2\pi)^{-1} {\rm ln}(R_0/r_{1,2})$
at separations $r_{1,2}\ll R_0$
 in the weak-coupling limit,
$\Lambda_0, \Lambda_1\rightarrow\infty$ (cf. ref. \cite{bulaevskii}).  
A scale transformation
$\vec r_m = l_{\phi}\cdot \vec x_m$
of the $2n$-dimensional integral above (\ref{auto3}) yields the final result
\begin{equation}
\overline{e^{i\theta_{l, l+n}}} \sim 
[(l_{\phi}/\Lambda_1)^2 {\rm ln}(\Lambda_1/l_{\phi})^2]^n
\label{result3}
\end{equation}
for  the asymptotic 
correlations of the superconducting order parameter across layers
near zero temperature.
The weakly coupled vortex-glass crosses over to
a 3D vortex lattice threaded by lines of dislocations
when the phase correlation length across layers, $L_{\phi}$,
exceeds the spacing between adjacent layers, $d$.
This crossover occurs at  a magnetic field
\begin{equation}
B_{D}(0) \sim  (l_{\phi}/a_{\rm vx})^2 (\Phi_0 / \Lambda_1^2)
\label{b_d}
\end{equation}
near zero temperature, 
at which point the argument between brackets on the
right-hand side of Eq. (\ref{result3}) is set to $1/e$.
The defective vortex lattice is decoupled across layers at
perpendicular magnetic
fields above $B_D$ (see Table \ref{mesa}),
where $\l_{\phi}\ll \Lambda_1$.

Consider again very weak Josephson coupling between adjacent layers,
such that $\l_{\phi}\ll \Lambda_1$.
Notice that this limit necessarily requires
high perpendicular magnetic fields compared to the 
naive decoupling scale,
$\Phi_0/\Lambda_1^2$,
by the inequality $a_{\rm vx} < l_{\phi}$.
Equation (\ref{result3})
then predicts short-range 
correlations of the superconducting order parameter across layers,
with a correlation length $L_{\phi}$ that is  less than the layer spacing $d$.
Imagine next that the 
quenched  disorder is reduced,
such that $\l_{\phi}\gg\Lambda_1$.
The argument in brackets on the right-hand side of 
Eq. (\ref{result3}) then notably does {\it not} diverge
towards positive infinity
with the ratio $l_{\phi}/\Lambda_1$
because of the 
logarithmic factor that originates from the 2D Greens function!
Instead, it attains a maximum value of order unity
at $l_{\phi}\sim \Lambda_1$.
Like in the cool-down from the vortex liquid, Eq. (\ref{result1}), 
these observations  indicate
that the correlation length $L_{\phi}$
for conventional superconducting
order across layers does not diverge at perpendicular magnetic fields
above the naive decoupling scale,
$B_{\perp} > \Phi_0/\Lambda_1^2$.
Unlike the case of vanishing thermal disorder ($\xi_{\phi}\rightarrow\infty$) in Eq. (\ref{result1}),
however,
the argument in brackets on the right-hand side of Eq. (\ref{result3})
diverges towards negative infinity
with vanishing quenched disorder ($l_{\phi}\rightarrow\infty$) 
because of the logarithmic factor.
That divergence  is  spurious.
The 2D Greens function (\ref{g2d}) is given by
$G^{(2)}(1,2) = (2\pi)^{-1} K_0 (r_{1,2}/R_0)$
in the limit $l_{\phi}\rightarrow\infty$, where 
${\rm cos}\, \theta_{l^{\prime}, l^{\prime}+1}^{(0)} = 1$.
Here, $K_0 (x)$ is a modified Bessel function, and 
$R_0 = \Lambda_1 / 2^{1/2}$.
Inspection of the original expression (\ref{auto3})
for the autocorrelator across layers
of the quenched superconducting order parameter
then yields 
the asymptotic result 
${\rm lim}_{n\rightarrow\infty}\,
 a_n (R_0 / \Lambda_1)^{2n} 
= [(1 + [1 + (\Lambda_1/\Lambda_0)^4]^{1/2}) / 4]^{n}$ 
for that quantity
as $l_{\phi}$ diverges.
Notice that the latter argument raised to the power $n$
instead saturates to a value that lies inside of the range $[0.5, 0.6]$,
which is  notably less than unity!
No evidence for conventional superconducting order of the vortex lattice
across a macroscopic number of layers therefore emerges from the above
perturbative analysis to lowest non-trivial order in the Josephson coupling
between layers,
at $B_{\perp} > \Phi_0 / \Lambda_1^2$.


\section{Discussion and Conclusions}
In conclusion, a duality analysis of the frustrated $XY$ model for the
mixed phase of layered superconductors with weak point defects finds
that  long-range vortex-glass order 
across layers
emerges out of  the vortex liquid
at weak Josephson coupling between layers.  
This is consistent with
recent Monte Carlo simulations of the same $XY$ model
that find evidence for a 
thermodynamic vortex glass phase\cite{nono1}\cite{olsson}.
It also potentially accounts for the recent  observation of a
thermodynamic vortex glass state in the mixed phase
of high-temperature superconductors that show extreme
layer anisotropy\cite{zeldov05}.
The analysis also indicates that 
the naive decoupling scale\cite{magnetic},
$\Phi_0/\Lambda_1^2$,
serves as an upper bound for the stability of the Bragg glass phase
as a function of perpendicular magnetic field
in the extreme type-II limit.
Previous theoretical  work on layered superconductors
predicts that the Bragg glass is stable 
to weak point pinning in general
at the extreme type-II limit\cite{K-V}.  
The discrepancy with
the present work is likely due to the use there of a 
criterion for the destruction of the 
Bragg glass phase that is too stringent.
In particular, the length $L_{\phi}$ along the field over which
the vortex lattice tilts by a lattice constant
is not divergent in ref. \cite{K-V}.
Also, the general robustness of the Bragg glass 
predicted by  ref. \cite{K-V}
at weak pinning 
conflicts with the belief that the
Bragg glass is generally {\it unstable} to invasion
by dislocations in  the limit of decoupled layers\cite{n-s},  
$\Lambda_1\rightarrow\infty$. 
A Bragg glass is also reported
at fields beyond
the naive decoupling scale   in ref. \cite{nono1},
where the
same $XY$ model
is  studied numerically by
Monte Carlo simulation.
The discrepancy with the stability bound established here
is likely due to a combination of finite-size effects
and of intrinsic pinning by the grid in each 2D $XY$ model (\ref{2dxy}).  
The last effect has been neglected here throughout.
Finally, Bragg peaks in neutron scattering that signal conventional
vortex-lattice order at long range have been observed in  the mixed phase
of extremely layered
high-temperature superconductors\cite{neutrons},
at fields below $500$ G.  That threshold is consistent with the
stability bound established here, $\Phi_0/\Lambda_1^2$,
if the Josephson penetration depth 
is bounded by $\Lambda_0 < 200$ nm.
Note that 
high layer anisotropy implies that the correction due to
magnetic screening ($\lambda_c$)
suggested by ref. \cite{magnetic}
can be ignored: $\Lambda_1\cong\Lambda_0$.


The two theoretical results just reviewed depend
critically on the existence of
a vortex-glass state for isolated layers in the vicinity of zero temperature.
Although recent experimental determinations of the current-voltage
characteristic in 2D arrays of Josephson junctions in weak
magnetic field obtain evidence for melting of the 2D vortex lattice
at transition temperatures $T_g^{(2D)}$ that are
in fact
 much greater than the
2D melting temperature of the pristine vortex lattice\cite{arrays},
theoretical arguments suggest that a perfectly conducting
vortex glass can exist only at zero temperature in two dimensions\cite{vinokur}.
Let us therefore consider the worst-case scenario,
$T_g^{(2D)} \rightarrow 0$.
The emergence of  long-range
vortex-glass order across layers from inside the weakly-coupled
vortex liquid  (\ref{result2}) survives this limit,
since   the 2D phase correlation length $\xi_{2D}$ remains divergent.
Secondly,
it is important to notice
that the field equation (\ref{fldeq}) used to obtain conventional
phase correlations across layers (\ref{result3}) inside of the vortex glass
is independent of the superfluid density $\rho_s^{(2D)}$.
This indicates that the stability bound in
perpendicular magnetic field
for the conventional vortex lattice,
$\Phi_0/\Lambda_1^2$,
survives the limit $T_g^{(2D)}\rightarrow 0$ as well.

\acknowledgments
The author thanks 
A. Koshelev for discussions. 
This work was performed under the auspices of the U.S. Department of Energy.

\begin{table}
\begin{center}
\begin{tabular}{|c|c|c|c|c|c|}
\hline
disorder index &
regime/phase &  $\overline{\langle{\rm cos}\, \phi_{l, l+1}\rangle}$
& $\rho_s^{\perp}/J_z$  & $L_{\phi} / d$ &  $\xi_{\perp} / d$  \\
\hline
\hline
$1$ & Bragg Glass & unity
& unity & $\infty$  & $\infty$ \\
\hline
$2$ & Defective Vortex Lattice 
& unity & unity & unity, or greater     & $\infty$  \\
$3$ & Vortex Glass & fraction
& fraction & fraction  & $\infty$ \\
\hline
$4$ & Critical Vortex Liquid &  fraction
& $0$ & fraction & unity, or greater \\
$5$ & Decoupled Vortex Liquid & fraction
& $0$  & fraction & fraction \\
\hline
\end{tabular}
\caption{Listed are
the conventional 
phase correlation length ($L_{\phi}$)
and the vortex-glass phase correlation length ($\xi_{\perp}$)
across equally spaced ($d$) layers,
as well as the corresponding
``cosine'' and
phase rigidity (see ref. \cite{jpr04a}),
for the various regimes found inside   the mixed phase of
an extremely type-II superconductor
at weak Josephson coupling between layers,
with weak point pinning.  
A   horizontal line marks a true phase transition.}
\label{mesa}
\end{center}
\end{table}


\begin{thebibliography}{}
\bibitem{integral}  The $2n$-dimensional integrals in 
Eqs. (\ref{auto1}) and (\ref{auto2}) 
that do not factorize
are achieved by
considering the 
respective displacements $\vec r_m - \vec r_{m+1}$
and $\vec R_m - \vec R_{m+1}$ as 
independent variables, 
and by imposing the constraint 
that they each sum to zero
through a $\delta$-function factor in the integrand (see ref. \cite{jpr00}).
Here, $\vec R_m$ is the center of mass of the inter-layer coordinates,
$\vec r_m$ and $\vec r_{\bar m}$, in Eq. (\ref{auto2}).
\bibitem{magnetic}
{Comparison  of Eq. (\ref{fldeq})
with the identical field equation
derived by Bulaevskii and Clem
in ref. \cite{B-C}
for the difference of the superconducting phase across adjacent layers 
suggests that the  effect of magnetic screening
absent in the frustrated  $XY$ model 
can be included simply by making the replacement
$2\Lambda_1^{-2} = 2\Lambda_0^{-2} +  \lambda_c^{-2}$,
where $\lambda_c$ denotes the London penetration depth associated
with Josephson supercurrents that flow across layers.}
\end{thebibliography}
\end{document}